\documentclass{emulateapj}
\usepackage{graphicx}
\usepackage{times}

\def\ltsima{$\; \buildrel < \over \sim \;$}
\def\simlt{\lower.5ex\hbox{\ltsima}}   
\def\gtsima{$\; \buildrel > \over \sim \;$}
\def\simgt{\lower.5ex\hbox{\gtsima}}

\def\msun{{\rm {M}}_{\odot}}
\def\etal{{\it et al.}}
\def\lsim{\mathrel{\lower .85ex\hbox{\rlap{$\sim$}\raise
.95ex\hbox{$<$} }}}
\def\gsim{\mathrel{\lower .80ex\hbox{\rlap{$\sim$}\raise
.90ex\hbox{$>$} }}}

\def \apjl  {ApJL}

\def \etal {et~al.~}
\def \chisq  {\ifmmode  \chi^2   \else  $\chi^2$  \fi}  
\def \chisqr {\ifmmode \chi^2_{\rm r} \else $\chi^2_{\rm r}$ \fi}
\def \spose#1{\hbox  to 0pt{#1\hss}}  
\def \lta{\mathrel{\spose{\lower 3pt\hbox{$\sim$}}\raise  2.0pt\hbox{$<$}}}
\def \gta{\mathrel{\spose{\lower  3pt\hbox{$\sim$}}\raise 2.0pt\hbox{$>$}}}
 
\def \ha  {\ifmmode H\alpha \else H$\alpha $ \fi}

\def \kms {\ifmmode  \,\rm km\,s^{-1} \else $\,\rm km\,s^{-1}  $ \fi }
\def \kpc {\ifmmode  {\rm kpc}  \else ${\rm  kpc}$ \fi  }  
\def \Msun {\ifmmode M_{\odot} \else $M_{\odot}$ \fi} 
\def \hMsun {\ifmmode h^{-1}\,\rm M_{\odot} \else $h^{-1}\,\rm M_{\odot}$ \fi}
\def \hhMsun {\ifmmode h^{-2}\,\rm M_{\odot}\else $h^{-2}\,\rm M_{\odot}$ \fi}
\def \Lsun {\ifmmode L_{\odot} \else $L_{\odot}$ \fi} 
\def \hhLsun {\ifmmode h^{-2}\,\rm L_{\odot} \else $h^{-2}\,\rm L_{\odot}$ \fi}

\def \LCDM {\ifmmode \Lambda{\rm CDM} \else $\Lambda{\rm CDM}$ \fi}
\def \lCDM {\ifmmode \Lambda{\rm CDM} \else $\Lambda{\rm CDM}$ \fi}
\def \lcdm {\ifmmode \Lambda{\rm CDM} \else $\Lambda{\rm CDM}$ \fi}
\def \sig8 {\ifmmode \sigma_8 \else $\sigma_8$ \fi} 
\def \OmegaM {\ifmmode \Omega_{\rm M} \else $\Omega_{\rm M}$ \fi} 
\def \OmegaL {\ifmmode \Omega_{\rm \Lambda} \else $\Omega_{\rm \Lambda}$\fi} 
\def \Deltavir {\ifmmode \Delta_{\rm vir} \else $\Delta_{\rm vir}$ \fi}

\def \rs {\ifmmode r_{\rm s} \else $r_{\rm s}$ \fi} 
\def \rrm2 {\ifmmode r_{-2} \else $r_{-2}$ \fi} 
\def \ccm2 {\ifmmode c_{-2} \else$c_{-2}$ \fi} 
\def \cvir {\ifmmode c_{\rm vir} \else $c_{\rm vir}$ \fi} 
\def \cbar {\ifmmode \overline{c} \else $\overline{c}$ \fi}

\def \R200 {\ifmmode R_{200} \else $R_{200}$ \fi} 
\def \Rvir {\ifmmode R_{\rm vir} \else $R_{\rm vir}$ \fi}

\def \v200 {\ifmmode V_{200} \else $V_{200}$ \fi} 
\def \Vvir {\ifmmode V_{\rm  vir} \else  $V_{\rm vir}$  \fi} 
\def  \Vhalo  {\ifmmode V_{\rm halo} \else $V_{\rm halo}$ \fi}

\def \M200 {\ifmmode M_{200} \else $M_{200}$ \fi} 
\def \Mvir {\ifmmode M_{\rm  vir} \else $M_{\rm  vir}$ \fi}  
\def \Mshell  {\ifmmode M_{\rm shell} \else $M_{\rm shell}$ \fi}

\def \Nvir {\ifmmode N_{\rm  vir} \else $N_{\rm  vir}$ \fi}  

\def \Jvir {\ifmmode J_{\rm vir} \else $J_{\rm vir}$ \fi} 
\def \Jshell {\ifmmode J_{\rm shell} \else $J_{\rm shell}$ \fi}

\def \Evir {\ifmmode E_{\rm vir} \else $E_{\rm vir}$ \fi} 

\def \lam {\ifmmode \lambda  \else $\lambda$ \fi} 
\def \lamp {\ifmmode \lambda^{\prime} \else $\lambda^{\prime}$  \fi} 
\def \lampc {\ifmmode \lambda^{\prime}_{\rm c} \else
  $\lambda^{\prime}_{\rm c}$  \fi} 
\def \lambar {\ifmmode \bar{\lambda}  \else  $\bar{\lambda}$  \fi}  
\def  \lampbar  {\ifmmode \bar{\lambda^{\prime}} \else
  $\bar{\lambda^{\prime}}$\fi} 
\def \siglam {\ifmmode \sigma_{\lambda} \else $\sigma_{\lambda}$ \fi} 
\def \siglamp {\ifmmode                \sigma_{\lambda^{\prime}} \else
$\sigma_{\lambda^{\prime}}$\fi}

\def \Rd {\ifmmode R_{\rm d} \else $R_{\rm d}$ \fi} 
\def \Rs {\ifmmode R_{\rm s} \else $R_{\rm s}$ \fi}  
\def \Rd {\ifmmode R_{\rm d} \else $R_{\rm d}$ \fi}  
\def \Rcool  {\ifmmode R_{\rm  cool}  \else $R_{\rm cool}$ \fi} 
\def \RIII {\ifmmode  3.2\Rs \else $3.2\Rs$ \fi} 
\def \RII {\ifmmode 2.2\Rs \else $2.2\Rs$  \fi} 
\def \Reff {\ifmmode R_{\rm eff} \else $R_{\rm  eff}$ \fi} 
\def  \rb {\ifmmode r_{\rm b}  \else $r_{\rm b}$ \fi}

\def  \Sigmacrit   {\ifmmode  \Sigma_{\rm  crit}   
\else  $\Sigma_{\rm crit}$\fi} 
\def \Sig0 {\ifmmode \Sigma_{0} \else $\Sigma_{0}$ \fi}

\def \muI {\ifmmode \mu_{0,I} \else $\mu_{0,I}$ \fi}

\def \mgal {\ifmmode m_{\rm gal} \else $m_{\rm gal}$ \fi} 
\def \md {\ifmmode m_{\rm d} \else $m_{\rm d}$ \fi} 
\def \ms {\ifmmode m_{\rm   s}   \else   $m_{\rm   s}$   \fi}   
\def   \mdbar   {\ifmmode {\overline{m}}_{\rm d} \else
  ${\overline{m}}_{\rm d}$ \fi} 
\def \msbar {\ifmmode  \bar{m}_{\rm  s}  \else  $\bar{m}_{\rm s}$
  \fi}  
\def  \Md {\ifmmode M_{\rm d}  \else $M_{\rm d}$ \fi} 
\def  \Ms {\ifmmode M_{\rm s} \else $M_{\rm  s}$ \fi} 
\def \Mb {\ifmmode  M_{\rm b} \else $M_{\rm b}$ \fi} 
\def \Mstar {\ifmmode  M_{\rm star} \else $M_{\rm star}$ \fi}
\def \Mdisc {\ifmmode M_{\rm disc} \else $M_{\rm disc}$ \fi}

\def \Jd {\ifmmode J_{\rm d} \else $J_{\rm d}$ \fi} 
\def \Jb {\ifmmode J_{\rm b} \else $J_{\rm b}$ \fi}  
\def \fb {\ifmmode  f_{\rm b} \else $f_{\rm b}$ \fi}

\def  \jd  {\ifmmode j_{\rm  d}  \else  $j_{\rm  d}$ \fi}  
\def  \jdmd {\ifmmode \frac{j_{\rm  d}}{m_{\rm d}} \else
  $\frac{j_{\rm d}}{m_{\rm d}}$ \fi} 
\def \fj {\ifmmode f_{\rm j} \else $f_{\rm j}$ \fi} 
\def \ft {\ifmmode f_{\rm t}  \else $f_{\rm t}$ \fi} 
\def  \fM {\ifmmode f_{\rm M} \else $f_{\rm M}$ \fi}

\def  \Vd {\ifmmode  V_{\rm  d}  \else $V_{\rm  d}$  \fi} 
\def  \Vcool {\ifmmode V_{\rm cool} \else $V_{\rm cool}$ \fi} 
\def \Vcirc {\ifmmode V_{\rm circ}  \else $V_{\rm circ}$  \fi} 
\def \VIII  {\ifmmode V_{3.2} \else $V_{3.2}$ \fi} 
\def  \VII {\ifmmode V_{2.2} \else $V_{2.2}$ \fi}
\def \Vobs {\ifmmode V_{\rm obs}  \else $V_{\rm obs}$ \fi} 
\def \Vdisc {\ifmmode V_{\rm disc} \else  $V_{\rm disc}$ \fi} 
\def \Vmax {\ifmmode V_{\rm  max} \else  $V_{\rm max}$  \fi} 
\def  \Vmaxobs{\ifmmode V_{\rm max}^{\rm obs}\else  $V_{\rm max}^{\rm
    obs}$\fi}  
\def \Vtot {\ifmmode V_{\rm tot} \else $V_{\rm tot}$  \fi} 
\def \Vrot {\ifmmode V_{\rm rot} \else  $V_{\rm rot}$  \fi} 
\def  \Vflat {\ifmmode  V_{\rm  flat} \else $V_{\rm flat}$ \fi}

\def \Ups {\ifmmode \Upsilon  \else $\Upsilon$ \fi} 
\def \YB {\ifmmode \Upsilon_B \else $\Upsilon_B$ \fi} 
\def \YI {\ifmmode  \Upsilon_I  \else $\Upsilon_I$ \fi} 
\def \DeltaIMF {\ifmmode \Delta_{\rm IMF} \else $\Delta_{\rm IMF}$ \fi}

\def\LCDM{$\Lambda$CDM }

\def\c200{$c_{200}$}

\shorttitle{}
\shortauthors{}

\begin{document}

\title{Small Dwarf Galaxies Within Larger Dwarfs: Why Some Are Luminous While Most Go Dark}

\author{Elena D'Onghia\altaffilmark{1,2} \& George Lake\altaffilmark{1}}
\affil{Institute for Theoretical Physik, 
 University of Zurich, Winterthurerstrasse 190, 8057 Zurich, Switzerland}
\email{elena@physik.unizh.ch}


\altaffiltext{2}{Marie Curie Fellow}

\begin{abstract}

We consider the possibility that the Magellanic Clouds were the
largest members of a group of dwarf galaxies that entered the Milky
Way (MW) halo at late times.  Seven of the eleven brightest satellites
of the MW may have been part of this system.  The proximity of some
dwarfs to the plane of the orbit of the Large Magellanic Cloud (LMC)
has been used to argue that they formed from tidal debris from the LMC
and Small Magellanic Cloud (SMC).  Instead, they may owe to the tidal
breakup of the Magellanic Group.  This can explain the association of
many of the dwarf galaxies in the Local Group with the LMC system.  It
provides a mechanism for lighting up dwarf galaxies and reproduces the
bright end of the cumulative circular velocity distribution of the
satellites in the MW without invoking a stripping scenario for the
subhalos to match the satellite distribution expected according to CDM
theory.  Finally, our model predicts that other isolated dwarfs will
be found to have companions.  Evidence for this prediction is provided
by nearby, recently discovered dwarf associations.

\end{abstract}


\section{Introduction}

Dwarf galaxies in the Local Group are puzzling for several reasons.
Some of them appear to be orbiting in roughly the same plane, and
observations suggest that this plane contains the Magellanic Clouds
along with some dwarf spheroidals.  This association of dwarfs with
the plane of the orbit of the LMC has been used to argue that dwarf
galaxies formed from tidal debris from the LMC and SMC (Kroupa \etal
2005).  However, this suggestion is challenged by the large dark
matter content of the dwarf galaxies, which is contrary to what is
expected if they are tidal debris (e.g. Barnes \& Hernquist 1992).

Another interesting aspect of the Local Group dwarf galaxies is that
simulations with only dark matter predict that the subhalos should
outnumber the modest population of observed, luminous dwarfs orbiting
the MW and M31 by a factor of 10 to 100 (Kauffman et al. 1993; Klypin
et al. 1999; Moore et al. 1999).  This discrepancy between the
predicted and observed numbers of dwarf galaxies has become known as
the {\it missing dwarf problem}. The population of ultra-faint dwarfs
around the MW and M31 found in the Sloan Digital Sky Survey (Willman
et al. 2005a,b; Zucker et al. 2006) has increased the number of
observed satellites by a factor of two.  It is unclear, however,
whether this can reconcile theory and observation (Simon \& Geha
2007).

Cosmological solutions to the missing dwarf problem include modifying
the power spectrum on small scales (e.g.  Zentner \& Bullock 2003, and
references therein) and changing the nature of the dark matter, by
assuming a warm dark matter particle (e.g. Colin et al. 2000;
Avila-Reese et al. 2001) or by invoking decay from a nonrelativistic
particle (Strigari et al. 2007).  Proposed astrophysical solutions
typically appeal to feedback effects associated with stellar evolution
or heating from UV radiation to inhibit the formation of dwarfs by
suppressing star formation in low mass halos (e.g., Bullock et
al. 2000; Somerville 2002; Benson et al. 2002).

An alternate scenario has been proposed by Kravtsov, Gnedin \& Klypin
(2004) in which the dwarf spheroidals we observe today were once much
more massive objects that have been reduced to their present mass by
tidal stripping.  On larger scales, the substructure function inside
clusters and groups is not well-determined and little is known about
the behavior of substructure over a range of masses (D'Onghia \& Lake
2004).
Data show a clear deficiency of substructure in systems of $\sim
10^{12} \msun$, characteristic of the mass of the MW, and mixed results
for systems with larger masses of a few $\times 10^{13} \msun$
(e.g. D'Onghia et al. 2008).

In this paper, we propose a scenario in which the Magellanic Clouds
and seven of the eleven dwarf galaxies around the MW were accreted as
a group of dwarfs which was disrupted in the halo of our Galaxy.  This
possibility is motivated by observations described in the next section
which indicate that dwarfs are often found in associations and
theoretically by numerical simulations (e.g. Li \& Helmi 2008) which
show that subhalos are often accreted in small groups.

If the LMC, SMC, and some dwarfs fell into the Milky Way in a group
rather than individually, they are more likely to end up orbiting in
roughly the same plane than according to earlier models.  For example,
Libeskind \etal (2005) determined that subhalos are anisotropically
distributed in cosmological CDM simulations and that the most massive
satellites tend to be aligned with filaments.  Zentner et al. (2005)
suggested that the accretion of satellites along filaments in a
triaxial potential leads to an anisotropic distribution of satellites.
Systems anisotropically distributed falling into the Galactic halo
are, however, unlikely to lie in a plane consistent with the orbital
and spatial distribution of the MW satellites (see e.g. Metz, Kroupa
\& Libeskind 2008).  As we argue further below, our scenario also
provides a new mechanism for lighting up the dSphs in the Local Group
that naturally reproduces the bright end of the cumulative circular
velocity distribution of the satellite dwarf galaxies in the MW.


\subsection{Evidence for a Disrupted Magellanic Group}

It has been recognized for many years that many dwarf galaxies may be
associated with the Magellanic Clouds (Lynden-Bell 1976, Fusi Pecci et
al. 1995; Kroupa \etal 2005).  The number of dwarfs possibly related
to the Magellanic Plane Group (Kunkel \& Demers 1976) has increased
with time and now includes the following {\it candidates}:
Sagittarius, Ursa Minor, Draco, Sextans and LeoII.  Of the dwarfs
known before the recent flurry of discoveries, 7 out of 10 within
$\sim 200$ kpc might well be connected with the Magellanic Clouds.
The remaining three -- Fornax, Sculptor and Carina -- have been
proposed to be part of a second grouping (Lynden-Bell 1982).  The
distribution of distant halo globular clusters has been used to
reinforce the existence of both groups (Kunkel 1979; Majewski 1994).

The LMC and SMC have long been viewed as a pair owing to their spatial
proximity.  They have been modeled as either being currently bound or
having only become unbound on a very recent perigalacticon passage
(Lin \& Lynden-Bell 1982).  The ``Magellanic Bridge" of neutral
hydrogen supports their being physically connected (Kerr \etal 1954).
New distance and proper motion measurements suggest that the clouds
are traveling together but have become unbound.  The LMC is at a
distance of $\approx 49$ kpc from the Galactic center and has a
tangential velocity of $\sim 345 \kms$ in galactic coordinates
(Kallivayalil \etal 2006a).  The proper motion of the SMC implies a
velocity relative to the LMC of $\sim 80 \kms$ at a separation of 23
kpc (Kallivayalil \etal 2006b).  Given the circular velocity of the
LMC ($76 \kms$) and the group's large initial virial radius ($\sim 75$
kpc), the relative velocity of the SMC appears modest.  However, the
current tidal radius of the LMC group is only $\sim 10$ kpc, so the
SMC is likely unbound, although it can travel with the LMC for a long
time owing to its relative retrograde motion and the larger effective
tidal radius for such an orbit (Read \etal 2006).  Simply put, the SMC
travels in an epicycle around the LMC so that they remain near for
several orbits after tidal breakup (Kallivayalil \etal 2006b).

The orbits of the pair depend on the total mass of the MW.  For a
traditional mass model with a total virial mass of $2 \times 10^{12}
\msun$, the orbit of the clouds currently has a rough peri- and
apogalacticon of 50 and 150 kpc (Kallivayalil \etal 2006b) 
respectively, with an implied apogalacticon 10 Gyr ago of $\sim 250$
kpc and little evolution in the perigalacticon, if dynamical friction
is included.  In this picture, the Magellanic group likely entered the
MW halo at a redshift between 2 and 3 or even at later times.  Recent
analysis by Besla et al. (2007) using a broader range of halo models
motivated by cosmological simulations admits the possibility that
the LMC has just fallen into the MW, is moving at close to the escape
velocity at its location, and is approaching its orbital pericenter
for the first time.

Here, we examine the possibility that the LMC, SMC, and those dwarfs
whose orbits are similar to those of the Magellanic Clouds were all
originally part of a group that was accreted by the Milky Way and
tidally disrupted.  This ``LMC group'' was dominated by the LMC and
had a parent halo circular velocity of $~75\kms$ (Kim et al. 1998)
with its brightest satellite, the SMC, having a rotation velocity of
$~60 \kms$ as estimated from its HI distribution (Stanimirovic et
al. 2004).  There is considerable evidence for tidal debris from the
LMC group, supporting the proposal that it was tidally disrupted.  For
example, microlensing studies of the Magellanic Clouds find more self
lensing than anticipated (Alcock \etal 1997, Palanque-Delabrouille
\etal 1998).  This was interpreted as self-lensing by the Magellanic
Clouds, although inconsistent with the structure of the LMC disk (Gyuk
\etal 2000).  Zaritsky and Lin (1997) find direct evidence for an {\it
intervening} stellar population which may comprise tidal debris.

An interesting characteristic of the LMC group is that its orbital
angular momentum is comparable to the entire MW, but is oriented at
90$^o$ relative to the disk plane.  The specific angular momentum of
an exponential disk rotating with constant velocity $v_{\rm{circ}}=220
\kms$ and scale length $r_s=2.8$ kpc is 2 $v_{\rm{circ}} r_s \sim 560$
kpc km s$^{-1}$.  The specific angular momentum of the LMC is its
galactocentric distance times its transverse velocity relative to the
galactic center.  From Kallivayalil \etal (2006b), this is $1.8 \times
10^4$ kpc km s$^{-1}$.  Hence, the LMC group has a specific angular
momentum that is roughly 20 times greater than the disk of the MW.
The mass of the LMC group was approximately $\sim 0.04$ the mass of
the MW.  If we assume that the entire halo of the galaxy has the same
specific angular momentum as the disk, the total angular momentum of
the LMC group is greater than the product of the entire mass of the MW
system times the specific angular momentum of the disk (e.g. Fich \&
Tremaine 1991; Besla et al. 2007).  This tilts the angular momentum
vector of the MW system by $\sim 45^o$ and may have significant
implications for comparing cosmological simulations of angular
momentum to present day galaxies (see e.g. Navarro, Abadi \& Steinmetz
2004).

\section{Evidence for associations of nearby dwarf groups}

In addition to the expectation that abundant populations of satellites
should orbit large galaxies, CDM theory predicts that many dwarf
galaxies exist in the field.  Numerical simulations indicate that the
mass function of subhalos should be nearly independent of the mass of
the parent halo (Moore \etal 1999).  Thus, groups of dwarf galaxies
are a natural outcome of CDM models on smaller mass scales.  However,
like low mass satellites, these systems are difficult to observe.

Tully (2006) discovered a number of associations of dwarf galaxies
within 5 Mpc of the MW.  These groups have properties expected for
bound systems with 1-10x10$^{11}$ M$_{\odot}$, but are not dense
enough to have virialized, and have little gas and few stars.  Of the
eight associations compiled by Tully (2006), there are only three for
which the two brightest galaxies differ by at least 1.5 magnitudes:
NGC3109, NGC1313 and NGC4214.  In the other five, the two brightest
galaxies are certain to merge if the associations collapse and
virialize.  We list the properties of the eight associations of dwarfs
in Table 1.  The other five low luminosity dwarf galaxies with
associated companions listed in Table 1 are: NGC55, NGC784, ESO
154-0123, DDO190 and DDO47.

For each association, the largest dwarf galaxy is considered.  The
columns in Table 1 list the name of the nearby low luminosity galaxies
with associated companions, the circular velocity of the largest dwarf
galaxy in each association $v_{\rm{par}}$ (which is assumed to be the
parent halo circular velocity), the circular velocity of the smallest
dwarf galaxy in the association, and the minimum number of observed
members N$_{\rm{obs}}$.  Magnitudes of member galaxies are converted
to circular velocity assuming a Tully-Fisher relation in the B band.
Note that of the three associations with the two brightest galaxies
differing by at least 1.5 magnitudes, NGC 3109 has a rotation velocity
roughly equal to the LMC, while NGC 1313 and NGC4214 are larger with
rotation velocities of 88 and 131$\kms$ respectively.

For our analysis, we compare the cumulative circular velocity
distribution of each dwarf which is part of the association with the
one inferred for the MW and for a model where the Magellanic group has
been disrupted into the MW.  The MW data we employ to construct the
cumulative circular velocity distribution of dwarf galaxies includes
the newest dwarfs with a minimum $\sigma = 3.3\kms$ and a correction
for incomplete sky coverage (Simon \& Geha 2007).

Figure 1 (left panel) shows the cumulative circular velocity
distribution function inferred for the dwarfs inside the eight
associations discovered by Tully (2006), the Magellanic Group
disrupted in the Galactic halo, and the MW satellite galaxies (Simon
\& Geha 2007).  Figure 1 implies that the nearby associations of
dwarfs have a similar cumulative circular velocity distribution
function to the MW, suggesting that such associations may be the
progenitors of the brightest dwarf satellites in the MW.  Thus, if
these associations of dwarfs are accreted into larger galaxies, they
can populate the bright end of the cumulative circular velocity
distribution function of satellites.  In order to match the cumulative
satellite distribution expected in CDM models, Kravtsov, Gnedin \&
Klypin (2004) suggested that the dwarf spheroidals we observe today
were once much more massive objects that have been reduced to their
present mass by tidal stripping.  In this way, the predicted high mass
end of the cumulative subhalo function can be matched. Figure 1
indicates that if the LMC group fell into the MW, the bright end of
the satellite galaxy distribution can be reproduced without invoking
any tidal stripping mechanism.

\section{LMC Group in a Cosmological Simulation}

To investigate the plausibility of our model, we examined a catalog of
high resolution galaxies in a cosmologically simulated volume to
identify an analog to an LMC group falling late into a MW galaxy.  The
simulation we employed was performed in a box 90 Mpc (comoving) on a
side with 300$^3$ particles, using the tree code PKDGRAV (Stadel
2001).  The cosmological parameters were chosen to match the WMAP3
constraints (Spergel \etal 2007): a present-day matter density,
$\Omega_0=0.238$; a cosmological constant, $\Omega_{\Lambda}$=0.762; a
Hubble parameter $h=0.73$ ($H_0=100\, h$ km s$^{-1}$ Mpc$^{-1}$); a
mass perturbation spectrum with spectral index, $n=0.951$; and a
normalization, $\sigma_8=0.75$.

In this volume, we located a loose group with properties similar to
the Local Group and which had an LMC type object being accreted by a
MW size halo at a relatively low redshift.  This loose group was
selected and resimulated at higher resolution using GRAFIC 2
(Bertschinger 2001).  There were almost 6 million particles inside the
high resolution region.  The subhalos at z=0 were identified using
SKID (Stadel 2001).

Figure 2 shows the group of dwarfs (at the bottom the figure) falling
into the MW halo at z=1.12.
The object that we refer to as the MW has a peak circular velocity of
$206 \kms$.  Within the virial radius of the simulated MW, there are a
total of 70 satellites with circular velocities greater than 10$\kms$.
There are 23 surviving satellites of the simulated Magellanic group
with 13 of them within the virial radius of the larger galaxy and 10
of them outside.

We conjecture that dwarf galaxies form in two extreme situations.  In
our model, dwarf galaxies accreted in LMC-like groups will be
luminous, whereas if they are not accreted in groups, they will be
dark, owing to the nature of the gas physics.  When gas is blown out
of a subhalo it eventually thermalizes to the virial temperature of
the parent halo, which is $2-5 \times 10^6$K for bright galaxies such
as the MW.  At this temperature, the cooling times are long enough
that there can be a considerable reservoir of hot gas and a subhalo
with a velocity scale of 10-30 $\kms$ will not reaccrete any gas, and
it will be dark.  However, in a small parent halo like the LMC, the
virial temperature is only $2 \times 10^5$K.  This is at the peak of
the cooling curve and the gas can therefore cool rapidly to $10^4K$.
In a group with this virial temperature, it is not possible to
maintain a gaseous halo capable of stripping a subhalo by ram
pressure.  The internal velocity scale of $30 \kms$ in the dwarf halos
might well be sufficient to reaccrete some of this gas and such
objects will become luminous dwarf galaxies.  (Note that simulations
by Keres et al. (2005) and analytical arguments (Birnboim \& Dekel
2003) show that gas in halos the size of the LMC will never be heated
to the virial temperature, but will instead be accreted in cold
filaments of $10^4K$. Indeed, this might be an additional mechanism
for satellites to re-accrete the gas.)

We display in Figure 1 (right panel) the cumulative peak circular
velocity distribution of the satellites contributed by the simulated
infalling group of dwarfs measured at z=0 within the virial radius of
the MW.  This is compared to the corresponding quantity for dwarfs
(filled squared symbols) in the MW which may have been part of an
accreted group: LMC, SMC, Sagittarius, Ursa Minor, Draco, Sextans and
Leo II (Lynden-Bell 1976; Kroupa \etal 2005).  In Figure 1, only
satellites which are accreted as part of the disrupted LMC group are
displayed, because those are the dwarf galaxies which light up in our
model according to the gas behavior described above.  The remainder of
the satellites in the simulations which are not accreted in groups but
are located at z=0 within the virial radius of the MW are assumed to
be dark in our scenario.

 
\section{Discussions and Conclusions}

We propose a model in which the LMC was the largest galaxy of a group
of galaxies that was accreted into the MW halo.  Our theory addresses
a number of outstanding problems in galaxy formation, particularly
those associated with dwarf galaxies, while making clear predictions
that can be tested in the near future.

Our picture can account for the apparent association of many of the
dwarf galaxies in the Local Group with the LMC system.  A scenario in
which dwarf galaxies are accreted in groups of dwarfs can explain the
planar orbital configuration populated by some dSphs in the MW.  Note
that planar structures are observed not only in the MW halo but the
dwarfs in the halo of M31 have also been grouped into planes as might
be expected if they entered in association with galaxies such as M33
(Koch \& Grebel 2006).  If satellites are distributed anisotropically
but are accreted individually, it is unlikely they would orbit in thin
planes.  However if they are accreted in groups they will eventually
orbit in planar structures.

We find that the LMC group can naturally reproduce the bright end of
the cumulative circular velocity distribution of the satellite
galaxies observed in the MW supporting a ``gas physics'' solution to
the missing satellite problem rather than one that proposes altering
the initial power spectrum.  The internal velocity scale of $30 \kms$
in the dwarf halos might well be sufficient to reaccrete some of this
gas leading to delayed episodes of star formation, a puzzling
phenomenon seen in the Local Group dwarfs (Mateo 1998, Grebel 1997).
Our model predicts that other isolated dwarfs will be found to have
companions down to this mass limit.  The recent discovery of Leo V
(Belokurov et al. 2008), a dwarf spheroidal companion of Leo IV and
the nearby dwarf associations supports our hypothesis.

\acknowledgments
ED is supported by a EU Marie Curie fellowship under contract MEIF-041569.


\clearpage

\begin{deluxetable}{rrrr}
\tablecolumns{6}
\tablewidth{0pc}
\tablecaption{Nearby Dwarf Galaxies Associations}
\tablehead{
\colhead{Name} & \colhead{V$_{\rm{par}}$ [km s$^{-1}$]} &  \colhead{V$_{\rm{min}}$ [km s$^{-1}$]} & \colhead{N$_{\rm{obs}}$}}  
\startdata
\tableline
NGC 1313   &   $131$ &   $26$  & $2$   \\
NGC 4214   &   $88 $ &   $26 $ & $6$   \\
NGC 3109   &   $75 $ &   $14 $ & $3$   \\
NGC 55      &   $104$ &   $21$  & $5$  \\
NGC 784     &   $77 $ &   $23$  & $4$  \\
ESO 154-023 &   $70 $ &   $61$  & $2$  \\
DDO 190     &   $67 $ &   $43$  & $3$  \\
DDO 47      &   $44 $ &   $29$  & $3$  \\
\tableline
\enddata
\end{deluxetable}  

\clearpage

\begin{figure}
\plotone{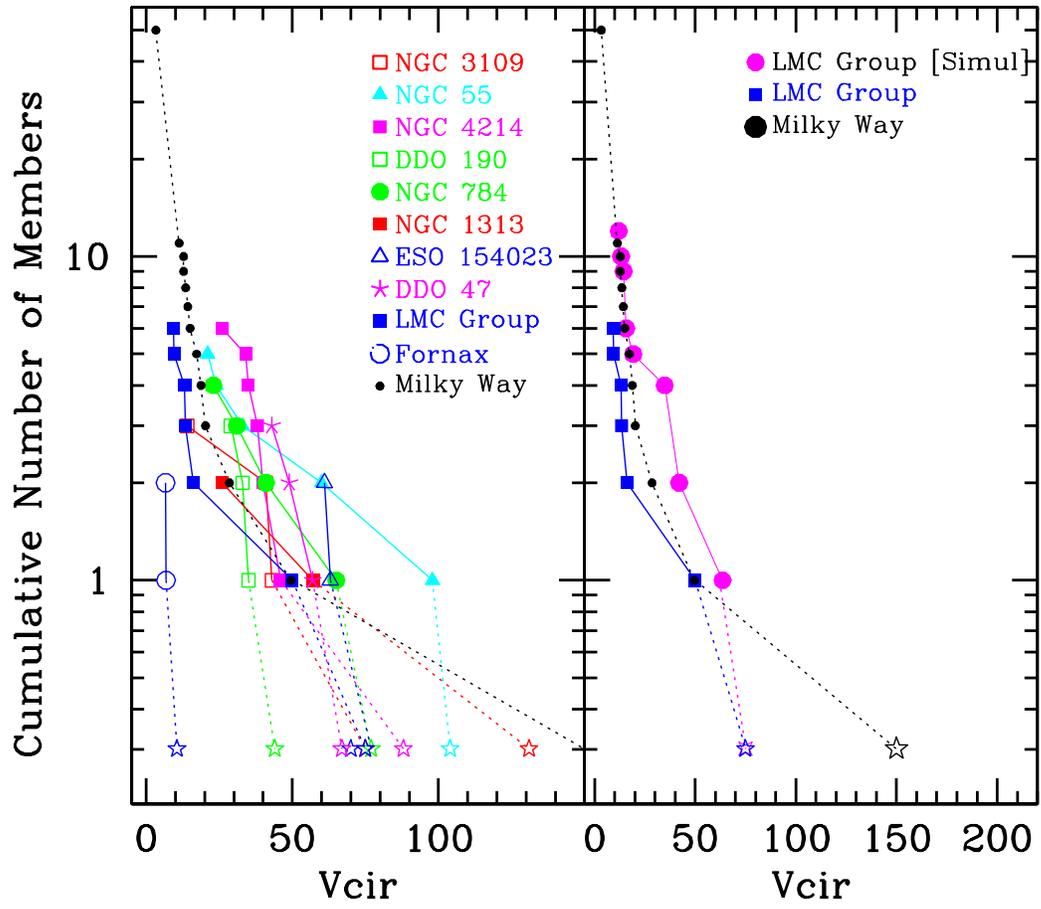}
\caption{Cumulative circular velocity distribution of the satellites of the LMC group as compared to the 
nearby dwarf associations (left panel) and to the simulated LMC group in a $\Lambda$CDM model (right panel).}
\end{figure}

\clearpage

\begin{figure}
\plotone{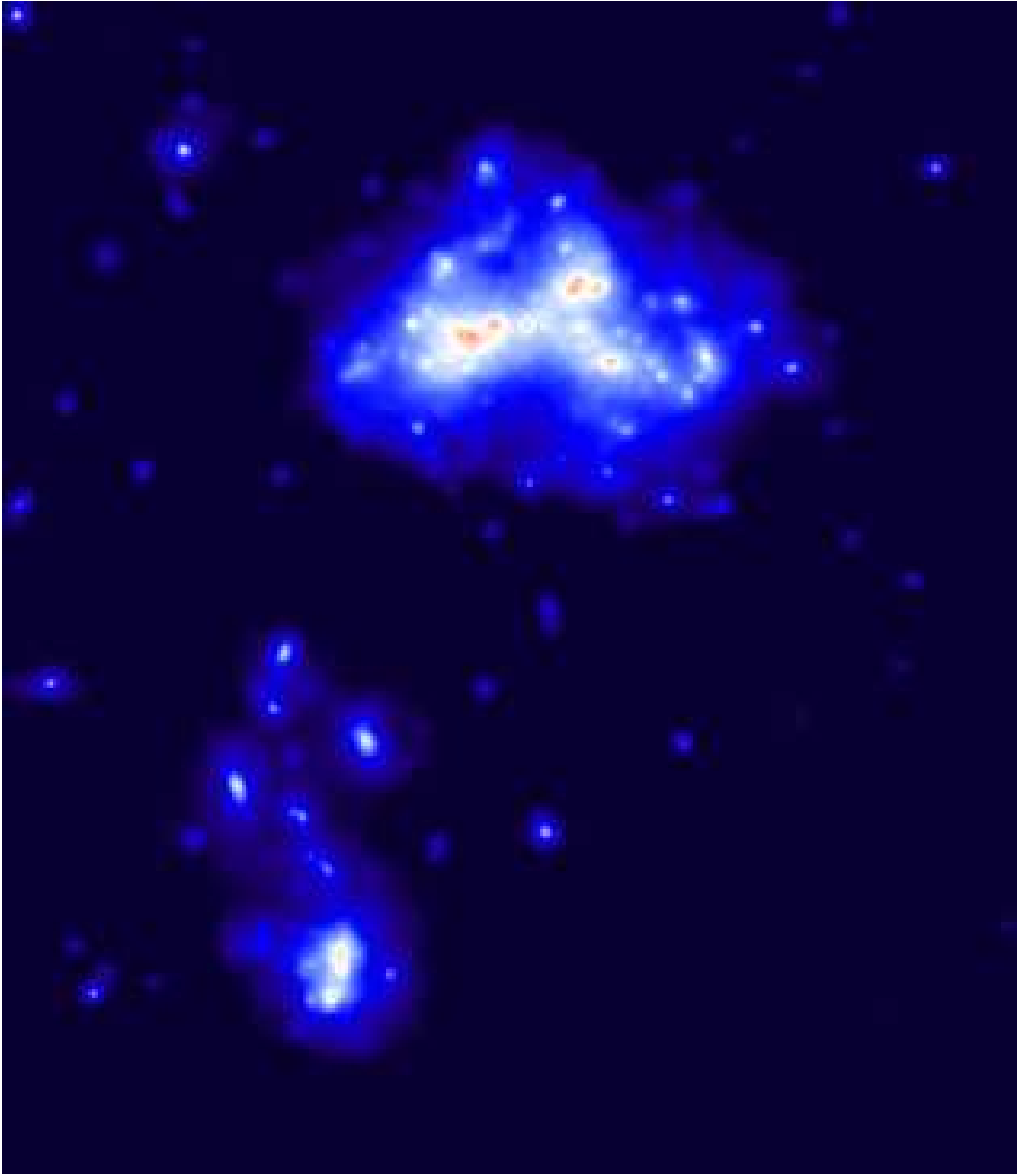}
\caption{Group of dwarfs dominated by the LMC (at the bottom) 
approaching the MW halo at $z>1$.}
\end{figure}

\end{document}